# Carrier-envelope phase effects on the strong-field photoemission of electrons from metallic nanostructures

**Björn Piglosiewicz[1,2], Slawa Schmidt[1,2], Doo Jae Park[1,2], Jan Vogelsang[1,2], Petra Groß[1,2], Cristian Manzoni[3], Paolo Farinello[3], Giulio Cerullo[3], and Christoph Lienau[1,2*]**

[1] *Institut für Physik, Carl von Ossietzky Universität, 26129 Oldenburg, Germany*
[2] *Center of Interface Science, Carl von Ossietzky Universität, 26129 Oldenburg, Germany*
[3] *IFN-CNR, Dipartimento di Fisica, Politecnico di Milano, 20133 Milano, Italy*
* Email: christoph.lienau@uni-oldenburg.de

**Abstract**

Sharp metallic nanotapers irradiated with few-cycle laser pulses are emerging as a source of highly confined coherent electron wavepackets with attosecond duration and strong directivity [1-6]. The possibility to steer, control or switch such electron wavepackets by light [7] is expected to pave the way towards direct visualization of nanoplasmonic field dynamics [8-11] and real-time probing of electron motion [12,13] in solid state nanostructures [14,15]. Such pulses can be generated by strong-field induced tunneling and acceleration of electrons in the near-field of sharp gold tapers within one half-cycle of the driving laser field [1,4,5]. Here, we show the effect of the carrier-envelope phase of the laser field on the generation and motion of strong-field emitted electrons from such tips. This is a step forward towards controlling the coherent electron motion in and around metallic nanostructures on ultrashort length and time scales.

**Introduction**

During the last two decades, strong-field phenomena in the interaction of light with atomic and molecular systems have led to a wealth of new physical phenomena, including the emission of high harmonic radiation [16,17] and the generation of attosecond light and X-ray pulses [18]. Key to those phenomena is a field-induced periodic modulation of the tunneling barrier, the acceleration of the photoemitted electrons within an essentially spatially homogeneous laser field and their recollision with the nuclei [19,20]. This has opened up the possibility to image molecular wavefunctions [12] and to probe electronic motion in real time. Attosecond science is enabled by the capability to control the carrier-envelope phase (CEP) of the light pulse, giving access to the electric field waveform rather than the intensity profile

More recently, related strong-field photoemission phenomena could be demonstrated for metallic nanostructures [4-6,21,22]. For sharply etched metallic tapers it could be shown that the emission process changes for sufficiently strong laser fields from multiphoton ionization (MPI) [1,2] towards above threshold ionization (ATI) [6,22] and optical field-induced tunneling [4,5,21]. Due to the large local field enhancement near the taper apex, the electron dynamics in the strong-field regime are substantially different from those in atomic and molecular systems. The corresponding pronounced spatial field gradients lead to a highly directional acceleration of the field-emitted electrons [5], escaping the local near field within less than half an optical cycle [4,5].

Here, we induce strong-field photoemission from ultrasharp nanometer-sized gold tapers with few-cycle near-IR pulses and analyze the effect of their CEP on the electron kinetic energy spectra. We observe a sensitive dependence of the electron acceleration dynamics on the CEP. Control of electron motion by the phase of a strong light field, which is of key importance for coherent control of HHG [17,23,24] and attosecond streaking [7,25,26] in gases, is here extended to solid-state nanostructures.

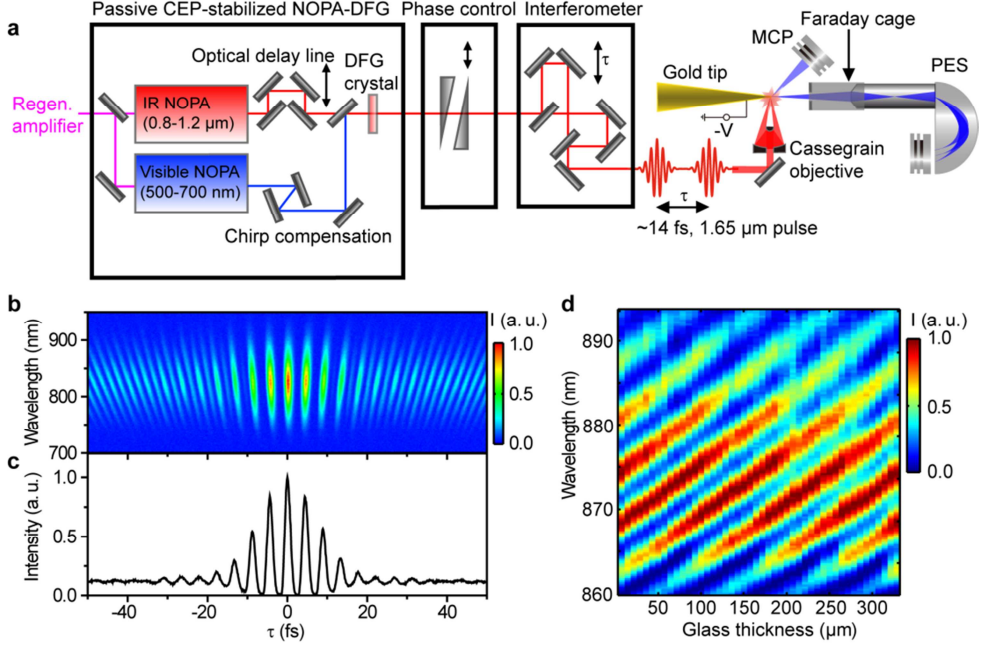

**Fig. 1. Electron generation from sharp metal tips with few-cycle CEP-stabilized pulses. a,** Generation of passively CEP-stabilized few-cycle near-IR pulses by difference frequency generation of pulses from two independently tunable noncollinear optical parametric amplifiers (NOPAs). CEP variation is achieved by wedge insertion, and a Michelson interferometer is used for peak intensity control and time-resolved measurements. The pulses are focused onto a sharply etched nanometer-sized gold tip to induce strong-field photoemission. The electron yield is measured by a microchannel plate (MCP) and kinetic energy spectra are recorded by a photoelectron spectrometer (PES). **b,** Frequency-resolved and **c,** spectrally integrated interferometric autocorrelation traces of the near-IR NOPA pulses centered at 1.65 µm, indicating a pulse duration of 14 fs. **d,** Measurement of the CEP variation in an $f$-to-$2f$ interferometer.

**Experimental**

We study photoemission from sharply etched, single crystalline gold tapers [27]. These tapers have a radius of curvature of down to 5 nm and are characterized by a large optical field enhancement factor of ~9 and a short decay length of the local near field at the taper apex of ~2 nm [5]. Few-cycle near-IR pulses are generated by difference frequency mixing of the output from two independently tunable noncollinear optical parametric amplifiers (NOPAs) [28]. The setup provides passively CEP-stabilized pulses with a center wavelength tunable between 1.3 and 1.9 µm. Their duration of 14 fs, corresponding to 2.6 cycles at 1.65 µm, is measured by interferometric frequency-resolved second-harmonic autocorrelation (Fig. 1b,c). The pulses display a high CEP stability with fluctuations of 66 mrad (root-mean-square) over a 600-s interval. The pulses, with their CEP controlled via a pair of fused silica wedges (Fig. 1d), are tightly focused to ~2 µm diameter onto the gold tips by an all-reflective objective [29]. We first investigate the effect of the laser intensity on the electron yield using incident pulses of up to 700 pJ energy (Fig. 2a). Assuming a field enhancement factor of $f = 9$ [5], the peak intensity at the tip is then varied from 0 to $f^2 \cdot I = 74$ TW/cm$^2$, corresponding to a local peak field strength $f \cdot E_0$ at the tip apex of up to 23.6 V/nm. At low intensities we observe a steep nonlinear increase in yield which is characteristic for MPI [2]. The order of nonlinearity of ~ 7.5 matches well with an estimated work function of gold of $\Phi = 5.5$ eV and an average photon energy of 0.7 eV ($\lambda = 1.65$ μm). At higher intensities above 15 TW/cm$^2$, the order of nonlinearity begins to decrease and a saturation of the yield is seen. This is the signature of the transition from MPI into a strong-field emission regime [21]. This transition appears for Keldysh parameters $\gamma = \dfrac{\sqrt{\Phi}}{\sqrt{2U_p}}$ of around 0.9. Here $U_p$ is the ponderomotive energy $U_p = \dfrac{(efE_0)^2}{4m\omega^2}$. The transition is accompanied by a distinct change of the corresponding photoelectron spectra. Narrow emission spectra characteristic for MPI broaden considerably with increasing intensity until we see, at the highest intensities, an essentially plateau-like kinetic energy spectrum (Figs. 2b-d). This broadening reflects the strong-field acceleration of photoemitted electrons within the steep near-field gradient at the tip apex.

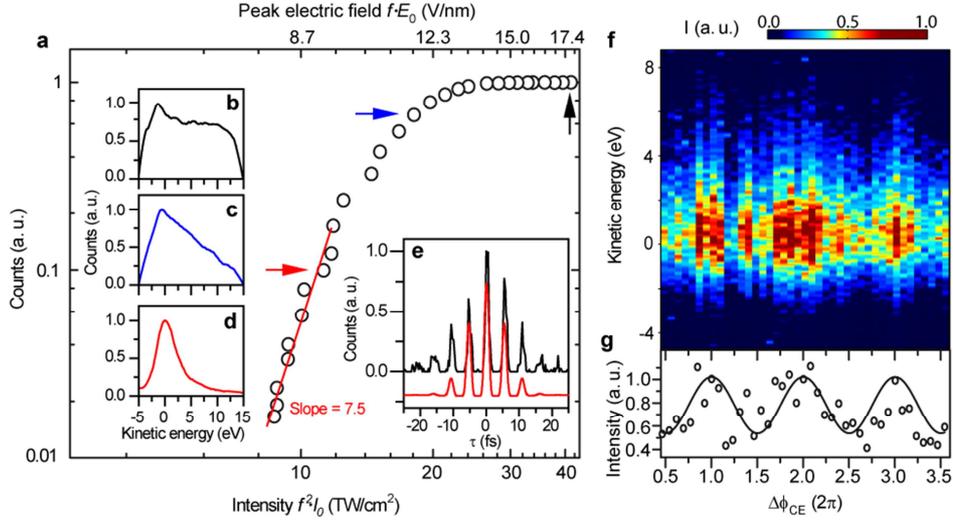

**Fig. 2. Transition from multiphoton to strong-field photoemission from sharp gold tips. a,** Electron yield as a function of intensity (bottom axis) and the corresponding peak electric field $f \cdot E_0$ (top axis) at the tip apex. For $f \cdot E_0 < 10$ V/nm, multiphoton ionization (MPI) dominates with a high nonlinearity of order ~7.5. For larger amplitudes, electron emission begins to saturate and is governed by optical-field-induced tunneling. Corresponding kinetic energy spectra **b-d** showing a transition to a broad plateau-like spectrum in the strong-field regime. **e,** Interferometric autocorrelation trace recorded by electron detection (black line) and simulation for a 16-fs pulse at 1.65 μm center wavelength with $f \cdot E_0 = 13$ V/nm (red line). **f,** Effect of the CEP variation on kinetic energy spectra, recorded with 14-fs pulses at 1.6 μm in the intermediate regime with $f \cdot E_0 = 11$ V/nm. Clearly pronounced variations in the electron yield as well as the width of the spectra are observed. **g,** Electron yield as a function of the CEP variation.

Simulations of the energy spectra within a modified two-dimensional Simpleman model [5] show that the energetic width of the plateau-like spectra at the highest is roughly given by 1.2 $U_p$. This provides a measure of the local field strength at the tip apex and, by comparison to the incident pulse intensity, an estimate of the field enhancement of $f \approx 9$. In these simulations, we assume that electrons are generated only during negative cycles, $E(t) < 0$, which accelerate the electrons away from the tip. We take an analytical model for the optical near field distribution around the apex. The local generation yield is deduced from a Fowler-Nordheim equation and the ejected electrons are accelerated as classical particles within the temporally oscillating and spatially varying near field. Possible recollisions with the tip are treated as perfectly elastic reflections with a yield of unity. We find that the energetic shape of the spectra depends sensitively on the decay length of the local near field. To approximate such plateau-like spectra we need to assume a short decay length of 2 nm for our sharp gold tapers.

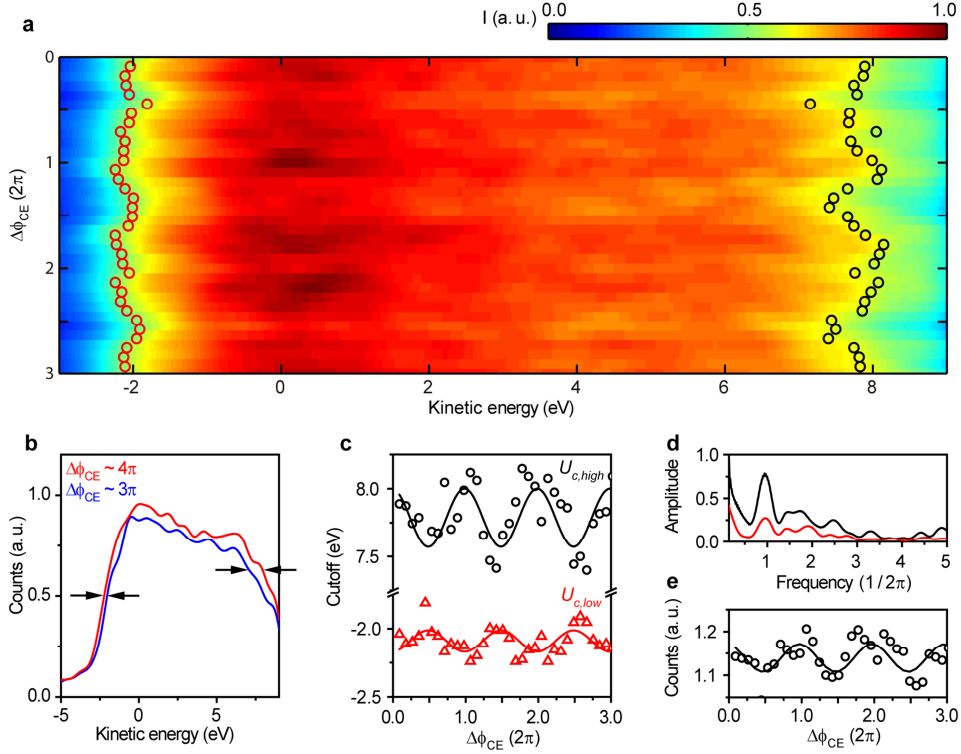

**Fig. 3: Experimental observation of CEP effects in strong-field electron emission from sharp metal tips. a,** Kinetic energy spectra recorded with 16-fs pulses at 1.65 μm with $f \cdot E_0 = 15.4$ V/nm while varying the CEP by wedge insertion. A clear modulation of the spectral width with a $2\pi$-periodicity is observed. The red and black circles indicate the low- and high-energy cutoffs, respectively. **b,** Two representative kinetic energy spectra recorded at $\Delta\Phi_{CE} \approx 3\pi$ (blue curve) and $4\pi$ (red curve) demonstrate the spectral variation when changing the CEP by $\pi$. Black arrows indicate the low- and the high-energy cutoff. **c,** Experimental high (black squares) and low (red triangles) cutoff energies extracted from **a** and sine functions fitted to these values. **d,** Fourier analysis of the experimental low- (red) and the high-energy (black) cutoff values from **c**. **e,** Total electron yield $I(\varphi_{CE}) = \int I(\varphi_{CE}, E_{kin}) dE_{kin}$ determined by integration over the spectra in **a**, showing the same periodicity as the cutoff energy variation.

Interferometric autocorrelation measurements, detecting the electron yield as a function of time delay between the incident pulses, show a few narrow emission spikes with a width of less than one femtosecond (Fig. 2e). The measurements agree well with simulations assuming 2.6 cycle pulses and electron generation during negative half cycles only. Higher electron yields measured for weaker field strengths are attributed to MPI neglected within the Fowler-Nordheim model. The simulations indicate electron emission during the central two cycles of the laser pulses.

Both the intensity and the spectral width of the kinetic energy spectra depend sensitively on the CEP of such few-cycle laser pulses. This is seen in Fig. 2f showing a series of 60 electron spectra recorded while varying the CEP by more than $6\pi$. The spectra are recorded at a local field strength of $f \cdot E_0 \approx 11$ V/nm, in the transition region between MPI and strong-field emission. We find variations of the total yield by about 50%. Interestingly, not only the high-energy but also the low-energy wing of the spectrum changes markedly with CEP. We find that this modulation results in a spectral width varying between 4 and 7 eV.

Similarly pronounced CEP effects are seen in the strong-field regime. CEP dependent spectra are recorded with 16-fs pulses with a higher local field amplitude of $f \cdot E_0 \approx$ 15 V/nm (Fig. 3a). In this regime we find that the effect of the CEP on the electron yield is much less pronounced, reflecting the saturation of the electron yield. Again, marked variations of the high- and low-energy cutoff of the plateau-like spectra are observed. Comparing spectra for $\Delta\Phi_{CEP} \approx 3\pi$ and $\Delta\Phi_{CEP} \approx 4\pi$, we see a variation of the high-energy cutoff by more than 1 eV (Fig. 3b). The variation of the low-energy cutoff is less pronounced and amounts to 0.3 eV. Plotting both quantities as a function of the CEP variation, clear out-of-phase oscillations and thus a narrowing and broadening of the spectrum with a period of $2\pi$, i.e., the oscillation period of the laser field, can be seen (Fig. 3c). The periodicity is confirmed by Fourier transforms (Fig. 3d) of the quantities plotted in 3c. Also the total electron yield oscillates with the laser period, with a reduced modulation contrast of 10%. The measurements present the first observation of CEP effects on the optical field emission and near field acceleration of electrons from an individual metallic nanostructure. Their generation and acceleration mechanism is distinctly different from above-threshold photoemission studied earlier [6].

**Numerical Simulation**

To account for these observations, we make use of the modified Simpleman model introduced above and study the effect of the CEP of a 2.6-cycle pulse at 1.65 μm (Fig. 4a) on the electron generation and motion. We describe the local near field in the vicinity of the tip by $f \approx 9$ and a decay length of 2 nm, values deduced experimentally. Of key importance are the acceleration of optical field-emitted electrons within spatial near-field gradients at the tip apex, and their dependence of the CEP and the birth time of the electron $\Delta t_B = t_B - t_0$ with respect to the negative amplitude maximum of the driving laser field at $t_0$. This acceleration is particularly pronounced for electrons emitted during the main cycle with maximum negative field amplitude.

Electrons with terminal kinetic energies close to the high-energy cutoff are emitted with birth times $-\frac{\pi}{2\omega} < \Delta t_B < 0$. Representative trajectories for $\Delta t_B = -\frac{\pi}{2\omega}$ are shown in Fig. 4b. Evidently these sub-cycle electrons are essentially accelerated only during the main half cycle. They move along the field lines at the taper apex, escaping the near field region before being back-accelerated during the subsequent positive half-cycle. Hence, their velocity remains constant at later times and their terminal kinetic energy is given by the near-field acceleration during the initial half-cycle. A CEP variation affects the local field amplitude at the apex and hence changes the terminal kinetic energy. The variation of the high-energy cutoff therefore is mainly due to strong-field acceleration of sub-cycle electrons within the spatial near field gradient. The terminal kinetic energy of these sub-cycle electrons decreases with increasing $\Delta t_B$ and together with the time-dependent generation yield explains the plateau-like energy spectra.

In case of a spatially homogenous driving field such as in conventional HHG, electrons with $E_{kin} \approx 0$ are generated near $\Delta t_B = 0$. Here the forward and backward acceleration during the positive and negative half-cycles effectively cancel out. Due to the large generation yield near $\Delta t_B = 0$ this results in the emission of a large number of slow electrons. CEP variations therefore mainly affect a small fraction of electrons with higher energy [30], while low-energy electrons are virtually unaffected. The strong near field gradient breaks the symmetry between forward and backward acceleration. Therefore the slowest electrons are released for slightly positive $\Delta t_B \approx \frac{0.1\pi}{\omega}$ (Fig. 4c). The net acceleration of these electrons is mainly gained during the second cycle after their emission. Thus their kinetic energy reflects the field amplitude during the subsequent negative half-cycle, explaining the CEP effect on the low energy electrons.

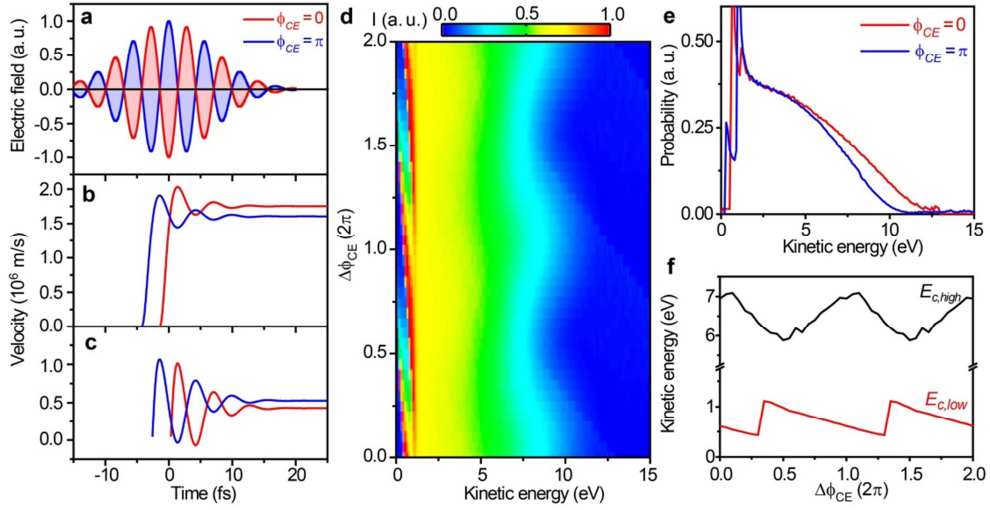

**Fig. 4: Numerical simulation of the CEP effect on strong-field photoemission. a,** Electric field of a 16-fs pulse centered at 1.65 μm for $\phi_{CE} = 0$ (red shaded area) and $\phi_{CE} = \pi$ (blue shaded area). **b,** Time evolution of the velocity of sub-cycle electrons emitted at the beginning of the main cycle, i.e., at $\Delta t_B = -\frac{\pi}{2\omega}$ for $\phi_{CE} = 0$ (red curve) and for $\phi_{CE} = \pi$ (blue curve). In both cases, the electrons are mainly accelerated during the initial half cycle and no quiver motion is seen. **c,** Velocity evolution for two electrons emitted from pulses with the same CEP but for $\Delta t_B = 0.1\frac{\pi}{\omega}$. Acceleration is mainly gained during the second cycle. **d,** Simulated kinetic energy spectra $I(\varphi_{CE}, E_{kin})$. **e,** Representative kinetic energy spectra, $I(0, E_{kin})$ and $I(\pi, E_{kin})$ shown as red and blue curves, respectively. **f,** High-energy cutoff extracted from **d** (black curve), following a sine function, and out-of-phase variation of the low-energy cutoff (red curve), resembling a saw-tooth function.

Simulations of CEP-dependent photoelectron spectra (Fig. 4d) confirm this scenario. They account for the plateau-like shape of the spectra (Fig. 4e), revealing the dominant contribution of sub-cycle electrons to the emission process. The marked periodic modulation of the high-energy cutoff (black curve in Fig. 4f) reflects the variation of the maximum field amplitude at the tip surface with CEP.

The CEP effect on the low-energy electrons is more involved. We see major contributions of electrons emitted during two subsequent negative half-cycles. Electrons emitted during the main cycle have a lower minimum kinetic energy since the field amplitude during the subsequent cycle is reduced. Electrons emitted earlier will experience a more pronounced acceleration by the main cycle field. This results in two discernible peaks in the low energy range (Fig. 4c). The effect of the CEP on the energy of these two peaks together with the time-dependent generation yield results in a sawtooth-like variation of the low-energy cutoff (red line in Fig. 4f), which is essentially out of phase with the high-energy cutoff modulation. The finite experimental energy resolution smears out the sawtooth-like modulation and leads to a periodic broadening and narrowing of the spectra when varying the CEP.

**Summary**

Taken together, our results present the first demonstration of CEP effects on the optical field emission of electrons from a single metallic nanostructure. The CEP controls the local amplitude of the near field during the generation cycle and therefore the electron acceleration within the near field gradient on a sub-cycle time scale. We believe that such field-driven control of the electron motion in the near field of solid state nanostructures can be seen as a new form of quantum electronics, paving the way towards the generation, measurement, and application of attosecond electron pulses.

**Methods**

**Tip fabrication.** The single-crystalline gold tapers were fabricated from polycrystalline gold wires (99.99%) with a diameter of 125 μm (commercially available from Advent Research Materials Ltd., Oxford, UK). After cleaning in ethanol, the wires are annealed at 800 °C for 8 h and then slowly, within another 8 h, cooled to room temperature. These annealed wires are then electrochemically etched in HCl (aq. 37%). For etching, rectangular voltage pulses with a frequency of 3 kHz and a duty cycle of 10% are applied between the wire and a platinum ring serving as the counter electrode. The tip shape is inspected by scanning electron microscopy.

**Experimental setup.** Spectrally tunable, few-cycle near-infrared pulses with a stable and controllable CEP are generated by difference frequency generation (DFG) of pulses from two noncollinear optical parametric amplifier (NOPA) stages [28]. Pulses from an amplified Titanium:Sapphire (Ti:Sa) regenerative amplifier (Newport, Spitfire Pro) with an energy of 0.5 mJ, 120-fs duration, 1-kHz repetition rate, and with a center wavelength of 800 nm are split into three parts to form the pump sources for white light (WL) generation and, after frequency doubling, for the two NOPA stages. WL is generated in a 2-mm thick sapphire plate and split into two parts, each part overlapping with a pump pulse in one the NOPA stages. Each stage consists of a 1-mm thick, type-I BBO crystal cut to sustain amplification from 540 nm to 650 nm, and from 870 nm to 890 nm, respectively. Both output pulses are collinearly aligned and temporally overlapped in a 0.3-mm thick, type-II cut BBO crystal to generate the difference frequency (DF), which results in passively CEP-stabilized pulses as short as 14 fs, tunable between $\lambda = 1.3$ and 1.9 μm, and with a pulse energy of up to 220 nJ. The CEP stability was checked in a conventional *f*-to-*2f* interferometer. The CEP of those pulses is controlled with a pair of fused silica wedges. A pulse pair with controllable inter-pulse delay is generated in a dispersion-balanced Michelson interferometer and focused onto the metal tips to a spot size of $0.6 \lambda / NA$ by an all-reflective objective [29] with a numerical aperture of $NA = 0.5$ under an angle of 90° with respect to the tip axis. Electrons emitted from the tip are monitored simultaneously by a microchannel plate (MCP) detector and a photoelectron spectrometer (PES, Specs Phoibos 100, aligned along the tip axis), to measure their yield and kinetic energy spectra, respectively.

**Data recording.** For measuring the electron yield, the MCP was used and laser intensity was varied by changing the arm length of the Michelson interferometer. The intensity scan was repeated 35 times, and the averaged curve is shown in Fig. 2a. For measuring CEP-dependent kinetic energy spectra, the PES detector was used and the CEP was controlled by wedge insertion. High-frequency noise from the NOPA was removed by using a zero-phase second order Butterworth filter with a cutoff frequency of 3.5·1/eV (Fig. 2f), and by applying a Fourier-filter with cutoff frequencies of 1.05·1/eV along the energy axis and of 3.5·1/(2π rad) along the phase axis (Fig. 3a). The values for the low- and high-cutoff energies (Fig. 3c) are extracted *before* applying the Fourier filter to the phase axis.

**Acknowledgments**

Financial support by the Deutsche Forschungsgemeinschaft (SPP1391), the European Union ("CRONOS") and the Korea Foundation for International Cooperation of Science and Technology (Global Research Laboratory project, K20815000003) is gratefully acknowledged. D.J.P. wishes to thank the Hanse-Wissenschaftskolleg for a personal fellowship.

**Author Contributions**

C.L. initiated the project. C.M., P.F., and G.C. designed and implemented the laser system. B.P., S.S., J.V., and D.J.P. implemented the setup and carried out the experiments. C.L., D.J.P., and P.G. developed the simulation model. D.J.P, P.G. and C.L. prepared the manuscript. All authors contributed to the final version of the manuscript.